\documentclass[preprintnumbers,amsmath,amssymb]{revtex4}

\usepackage{graphicx}
\usepackage{dcolumn}
\usepackage{bm}

\begin{document}

\title{\Large\bf  Quantum secure direct communication by Einstein-Podolsky-Rosen pairs and  entanglement swapping}

\author{Ting Gao$^{1,2,3}$,  Feng-Li Yan $^{3,4}$, Zhi-Xi Wang$^2$}

\affiliation {\footnotesize $^1$\sl College of Mathematics and Information Science,
Hebei Normal University, Shijiazhuang 050016, China\\
 $^2$ Department of Mathematics, Capital Normal University, Beijing 100037, China\\
$^3$ CCAST (World Laboratory), P.O. Box 8730, Beijing 100080, China\\
$^4$ College of Physics, Hebei Normal University, Shijiazhuang 050016, China} \maketitle

We present a quantum secure direct communication  scheme  achieved by swapping quantum entanglement. In this
scheme a set of ordered Einstein-Podolsky-Rosen (EPR) pairs is used as  a quantum information channel for
sending secret messages directly. After insuring the safety of the quantum channel, the sender Alice encodes the
secret messages directly by applying a series local operations on her particles sequence according to their
stipulation.
 Using three EPR pairs, three bits of secret classical information can
be faithfully transmitted from Alice to remote Bob without revealing any information to a potential
eavesdropper.
  By both Alice and Bob's
GHZ state measurement results, Bob is able to read out the encoded secret messages directly. The protocol is
completely secure if perfect quantum channel is used, because there is not a transmission of the qubits
 carrying the secret message between Alice and Bob in the public channel.\\


1. Introduction

 Since the first quantum cryptography protocol  using quantum mechanics
to distribute keys was proposed by Bennett and Brassard  in 1984 (called BB84) [1], numerous quantum
cryptographic protocols have been proposed, such as Ekert 1991 protocol (Ekert91)[2], Bennett-Brassard-Mermin
1992 protocol (BBM92) [3], B92 protocol [4] and others protocols [5-21].

Different from key distribution whose purpose is to establish a common random key between two parties, a secure
directly communication is to communicate important messages directly without first establishing a random key to
encrypt them.
  Recently, Shimizu and Imoto [22, 23] and Beige et al. [24] proposed  novel quantum secure direct
communication (QSDC) schemes, in which the two parties communicate important messages directly without first
establishing  a shared secret key to encrypt them and the message is deterministically sent through the quantum
channel, but can be read only after obtaining an additional classical information for each qubit. Bostr\"{o}m
and Felbinger [25] put forward a QSDC scheme, the "ping-pong protocol", which is insecure if it is operated in a
noisy quantum channel, as indicated  by W\'{o}jcik  [26]. Deng et al. [27] suggested a two-step quantum direct
communication protocol using Einstein-Podolsky-Rosen pair block. However, in all these QSDC schemes it is
necessary to send the qubits with secret messages  in the public channel. Therefore, Eve can attack the qubits
in transmission.

By using  Einstein-Podolsky-Rosen (EPR) pairs and teleportation  [28], Yan and Zhang [29] presented  a QSDC
scheme . By means of controlled quantum teleportation [30] we proposed two controlled QSDC protocols [31, 32],
one using Greenberger-Horne-Zeilinger (GHZ) states and teleportation, another with entangled states different
from GHZ states. Because there is not a transmission of the qubits carrying the secret messages between Alice
and Bob in the public channel, they are completely secure for direct secret communication as long as perfect
quantum channel is used.

Entanglement swapping [33] is a method that enables one to entangle two quantum systems that do not have direct
interaction with one another. In the virtue of entanglement swapping, we introduce a QSDC scheme  without
alternative measurements. In the scheme, the secure communication between two spatially separated parties (Alice
and Bob) is achieved via initially shared EPR pairs, which function as a quantum channel for faithful
transmission. After insuring the safety of the quantum channel, the sender Alice encodes the secret messages
directly by applying a series local operations on her particles sequence according to their stipulation. From
both Alice and Bob's GHZ state measurement results, which is communicated via classical channel, Bob is able to
read out the encoded secret messages directly. The secret messages are faithfully transmitted from a sender
(Alice) to a
remote receiver (Bob)  without revealing any information to a potential eavesdropper.\\

2. Quantum secure direct communication protocol by swapping quantum entanglement

 Suppose two distant parties, Alice and Bob, share three EPR pairs (three of the four Bell states)
 \begin{equation}\label{EPR}
|\Phi^{\pm}\rangle\equiv\frac{1}{\sqrt{2}}(|00\rangle\pm|11\rangle),
|\Psi^{\pm}\rangle\equiv\frac{1}{\sqrt{2}}(|01\rangle\pm|10\rangle).
\end{equation}
Without loss of generality, assume that Alice and Bob share  $|\Phi^+\rangle_{12}$,  $|\Psi^+\rangle_{34}$ and
$|\Phi^+\rangle_{56}$, where Alice has qubits 1, 3 and 5, and Bob possesses 2, 4 and 6. A measurement is
performed on particles 1, 3 and 5 with the GHZ basis,  $|P^{\pm}\rangle$, $|Q^{\pm}\rangle$, $|R^{\pm}\rangle$
and $|S^{\pm}\rangle$, then the total state $|\Phi^+\rangle_{12}\otimes|\Psi^+\rangle_{34}\otimes
|\Phi^+\rangle_{56}$ is projected onto $|R^+\rangle_{135}\otimes |P^+\rangle_{246}$, $|R^-\rangle_{135}\otimes
|P^-\rangle_{246}$, $|S^+\rangle_{135}\otimes |Q^+\rangle_{246}$, $|S^-\rangle_{135}\otimes |Q^-\rangle_{246}$,
$|P^+\rangle_{135}\otimes |R^+\rangle_{246}$, $|P^-\rangle_{135}\otimes |R^-\rangle_{246}$,
$|Q^+\rangle_{135}\otimes |S^+\rangle_{246}$ and $|Q^-\rangle_{135}\otimes |S^-\rangle_{246}$ with equal
probability of 1/8 for each. Here the GHZ basis are eight GHZ states
\begin{equation}\label{GHZ}
    |P^{\pm}\rangle\equiv\frac{1}{\sqrt{2}}(|000\rangle\pm|111\rangle),
    |Q^{\pm}\rangle\equiv\frac{1}{\sqrt{2}}(|001\rangle\pm|110\rangle),
    |R^{\pm}\rangle\equiv\frac{1}{\sqrt{2}}(|010\rangle\pm|101\rangle),
    |S^{\pm}\rangle\equiv\frac{1}{\sqrt{2}}(|011\rangle\pm|100\rangle).
\end{equation}
 Previous entanglement between qubits 1 and 2,  3 and 4, and 5 and 6 are now
swapped into entanglement of qubits 1, 3 and 5, and 2, 4 and 6. Although we considered entanglement swapping
with the initial state $|\Phi^+\rangle_{12}\otimes|\Psi^+\rangle_{34}\otimes|\Phi^+\rangle_{56}$, similar
results can be achieved with other Bell states. For example, when Alice and Bob originally share
$|\Psi^+\rangle_{12}$, $|\Phi^-\rangle_{34}$ and $|\Phi^+\rangle_{56}$, there are eight possible measurement
outcomes, $|S^-\rangle_{135}\otimes |P^+\rangle_{246}$, $|S^+\rangle_{135}\otimes |P^-\rangle_{246}$,
$|R^-\rangle_{135}\otimes|Q^+\rangle_{246}$, $|R^+\rangle_{135}\otimes|Q^-\rangle_{246}$,
$|Q^-\rangle_{135}\otimes|R^+\rangle_{246}$, $|Q^+\rangle_{135}\otimes|R^-\rangle_{246}$,
$|P^-\rangle_{135}\otimes|S^+\rangle_{246}$ and $|P^+\rangle_{135}\otimes|S^-\rangle_{246}$ with equal
probability 1/8.

Now two spatially separated  parties, Alice and Bob, wish to communicate in secret. In order to realize privacy,
the first step in our scheme is to establish quantum channel (EPR pairs). Obtaining these EPR pairs could have
come about in many different ways, for instance, one of Alice or Bob prepares a sequence of EPR pairs and then
send half of each pair to another.  Alice and  Bob  then choose  randomly a subset EPR pairs, and  do   some
appropriate tests of fidelity.  Passing the test certifies that
  they continue to hold sufficiently pure, entangled quantum states.  However,
 if tampering has occurred,
 Alice and Bob throw out the EPR pairs and start over.

 After insuring the security of the quantum channel (EPR pairs), Alice and Bob  begin  the second step of our
 scheme--- secure direct communication.  Both Alice and Bob divide all Bell states into $N$ ordered groups
 $\{\xi(1)_{12}, \eta(1)_{34}, \zeta(1)_{56}\}$, $\{\xi(2)_{12}, \eta(2)_{34}, \zeta(2)_{56}\}$, $\cdots$,
 $\{\xi(N)_{12}, \eta(N)_{34}, \zeta(N)_{56}\}$ at random,
 each group $\{\xi(i)_{12}, \eta(i)_{34}, \zeta(i)_{56}\}$
($i=1, 2, 3, \cdots, N$) include 3 Bell states. Particles 1, 3 and 5, and particles 2, 4 and 6  of each group
belong to Alice and Bob, respectively. Alice encodes information by local operations on EPR pairs. She can
perform on each of her particles 1 one of the following four unitary  operations
\begin{equation}\label{operation1}
 \sigma_{00}=I=|0\rangle\langle 0|+|1\rangle\langle1|,
~~\sigma_{01}=\sigma_x=|0\rangle\langle1|+|1\rangle\langle0|, ~~\sigma_{10}={\rm
i}\sigma_y=|0\rangle\langle1|-|1\rangle\langle0|, ~~ \sigma_{11}=\sigma_z=|0\rangle\langle0|-|1\rangle\langle1|,
\end{equation}
and on each of her particles 3 one of the two operations
\begin{equation}\label{operation2}
 \sigma_0=I=|0\rangle\langle 0|+|1\rangle\langle1|, ~~ \sigma_1=\sigma_x=|0\rangle\langle1|+|1\rangle\langle0|.
\end{equation}
 Alice and Bob assign three bits to Alice's operations as following encoding
\begin{equation}\label{m1}
 \sigma_{ij}\otimes\sigma_k\rightarrow ijk,  ~~ i, j, k=0, 1.
\end{equation}
Alice  applies  local operations  to each pair of her particles 1 and 3 according to the secret message
sequence. For instance, if the message to be transmitted is a sequence 110010101001011100, then she performs
local operations sequence $\sigma_{11}\otimes\sigma_0$, $\sigma_{01}\otimes\sigma_0$,
$\sigma_{10}\otimes\sigma_1$, $\sigma_{00}\otimes\sigma_1$, $\sigma_{01}\otimes\sigma_1$,
$\sigma_{10}\otimes\sigma_0$ on particles 1 and 3 of $\{\xi(1)_{12}, \eta(1)_{34}, \zeta(1)_{56}\}$,
$\{\xi(2)_{12}, \eta(2)_{34}, \zeta(2)_{56}\}$, $\cdots$, $\{\xi(6)_{12}, \eta(6)_{34}, \zeta(6)_{56}\}$,
respectively. Alice and Bob make GHZ measurements on particles 1, 3 and 5, and 2, 4 and 6, respectively. After
that Alice tells Bob her measurement results. According the outcomes of Alice's measurement, Bob can infer the
information that Alice transmits to her. The  specific steps of the QSDC scheme using entanglement swapping are
as follows:

(1) Alice  prepare EPR pairs and then sent half of each  to Bob, or vice versa. They both randomly divide all
Bell states into $N$ ordered groups $\{\xi(1)_{12}, \eta(1)_{34}, \zeta(1)_{56}\}$, $\{\xi(2)_{12},
\eta(2)_{34}, \zeta(2)_{56}\}$, $\cdots$,
 $\{\xi(N)_{12}, \eta(N)_{34}, \zeta(N)_{56}\}$, and denote $\xi(i)_{12}, \eta(i)_{34}, \zeta(i)_{56}$
 for three Bell states of Alice's particles 1, 3 and 5, and Bob's particles
  2, 4 and 6 in the  $i$-th group.

(2) Alice and Bob agree on each of local operations $\sigma_{ij}\otimes\sigma_k$ can carry three-qubit classical
information and encode $\sigma_{ij}\otimes\sigma_k$ as $ijk$, where $i, j, k=0, 1$.

(3) Alice encodes her messages on  EPR groups. Explicitly,  Alice applies a local operation on each pair of her
particles 1 and 3 according to the secret message sequence.

 Suppose Alice and Bob initially share Bell state $|\Phi^+\rangle_{12}, |\Phi^+\rangle_{34}, |\Phi^+\rangle_{56}$,
  then the originally
  total state of  them is
\begin{eqnarray}\label{original}
 |\Phi^+\rangle_{12}\otimes|\Phi^+\rangle_{34}\otimes|\Phi^+\rangle_{56}
 &=& \frac{1}{2\sqrt{2}}(|P^+\rangle_{135}\otimes|P^+\rangle_{246}+|P^-\rangle_{135}\otimes|P^-\rangle_{246}
 +|Q^+\rangle_{135}\otimes|Q^+\rangle_{246}
 +|Q^-\rangle_{135}\otimes|Q^-\rangle_{246}\nonumber\\
 &&+|R^+\rangle_{135}\otimes|R^+\rangle_{246}+|R^-\rangle_{135}\otimes|R^-\rangle_{246}
 +|S^+\rangle_{135}\otimes|S^+\rangle_{246}+|S^-\rangle_{135}\otimes|S^-\rangle_{246}).
\end{eqnarray}
If Alice wishes to transmit 111 to Bob,  then she performs a local operation $\sigma_{11}\otimes\sigma_1$ on
particles 1 and 3  and the
   state $\Phi^+_{12}\otimes|\Phi^+\rangle_{34}$ is turned into $|\Phi^-\rangle_{12}\otimes\Psi^+_{34}$.

(4) Alice makes a GHZ state measurement on her particles 1, 3 and 5.  Assume that her measurement outcome is
$|P^+\rangle_{135}$, then she can induce Bob's three particles 2, 4 and 6  in the state $|R^-\rangle_{246}$ by
the following equation:
\begin{eqnarray}\label{factual}
|\Phi^-\rangle_{12}\otimes|\Psi^+\rangle_{34}\otimes|\Phi^+\rangle_{56}
 &=& \frac{1}{2\sqrt{2}}(|R^-\rangle_{135}\otimes|P^+\rangle_{246}+|R^+\rangle_{135}\otimes|P^-\rangle_{246}
 +|S^-\rangle_{135}\otimes|Q^+\rangle_{246}
 +|S^+\rangle_{135}\otimes|Q^-\rangle_{246}\nonumber\\
 &&+|P^-\rangle_{135}\otimes|R^+\rangle_{246}+|P^+\rangle_{135}\otimes|R^-\rangle_{246}
 +|Q^-\rangle_{135}\otimes|S^+\rangle_{246}+|Q^+\rangle_{135}\otimes|S^-\rangle_{246}).
\end{eqnarray}

(5) Alice tells Bob that she has made a Bell measurement on her particles 1, 3 and 5 over a classical channel,
but does not mention the result of her measurement.

(6) Bob performs a GHZ-type measurement on his particles 2, 4 and 6, and  infers the outcome of Alice's
measurement.

 From the calculation of entanglement swapping (Eq.(\ref{original}))  and Bob's measurement outcome $|R^-\rangle_{246}$,
 Bob could determine  exactly that the outcome of Alice's measurement should be $|R^-\rangle_{135}$
 without Alice's local operation.

(7) Bob asks and gets Alice's measurement result publicly.

(8) Bob can read out Alice's secret message by comparing his calculation result  with Alice's practical
measurement outcome.

From the  measurement result $|P^+\rangle_{135}$ announced by Alice and his calculation result
$|R^-\rangle_{135}$, Bob can infer that Alice  has applied a local operation $\sigma_{11}\otimes\sigma_1$ on
particles 1 and 3, thus he obtains  Alice's message 111. Finally, the two distant parties have realized
deterministic secure direct communication.

Note: (A) The above protocol is also a quantum key distribution (QKD) scheme based on Bell states and
entanglement swapping, if Alice  applies her  local operations $\sigma_{ij}\otimes\sigma_k$, as determined by
$a$, a string of $3N$ random classical bits which she  creates on her  own. That is, depending on three random
classical bit $a$ which she generates, Alice  performs a local operation $\sigma_{ij}\otimes\sigma_k$ if $a=ijk$
($i, j, k=0, 1$) on  her particles 1 and 3. Alice and Bob agree upon in advance that each of the eight GHZ
states can carry three bits classical information and encode $|P^+\rangle$, $|P^-\rangle$, $|Q^+\rangle$,
$|Q^-\rangle$, $|R^+\rangle$, $|R^-\rangle$, $|S^+\rangle$, and $|S^-\rangle$  as 000, 001, 010, 011, 100, 101,
110 and 111, respectively. By Alice's measurement result $|P^+\rangle_{135}$, both Alice and Bob derive
$|R^-\rangle_{135} \xrightarrow{\sigma_{11}\otimes\sigma_1} |P^+\rangle_{135}$ and share three certain bits 111
and three random bits 101 in private. Therefore,
 in our proposed protocol, Alice  performs one local operation on her particles 1 and 3,
  Bob shares 3  certain bits and 3 random bits  with Alice secretly.

(B) (B1) Alice can also apply unitary operator
\begin{equation}\label{m21}
  I=|0\rangle\langle 0|+|1\rangle\langle1|,
 ~~{\rm i}\sigma_y=|0\rangle\langle1|-|1\rangle\langle0|
\end{equation} on particles 3,
 and she and Bob   agree beforehand as the following encoding:
\begin{equation}\label{m22}
  I\rightarrow 0,~~ {\rm i}\sigma_y\rightarrow 1,
\end{equation}
instead of that in the above protocol. (B2) In the above protocol, Alice   can  use local operations in
Eq.(\ref{operation2}) and Eq.(\ref{operation1}) on particles 1 and 3, respectively. Moreover, Alice and Bob
arrange the encoding as $\sigma_i\otimes\sigma_{jk}\rightarrow ijk$ for $i, j, k=0, 1$. (B3) The unitary
operations Eq.(\ref{operation1}) and Eq.(\ref{operation2}) performed by Alice  in above protocol can be replaced
by local operations in Eq.(\ref{m21}) and Eq.(\ref{operation1}), respectively. Furthermore,  Alice and Bob agree
on the decoding as $I\otimes\sigma_{jk}\rightarrow 0jk$, ${\rm i}\sigma_y\otimes\sigma_{jk}\rightarrow 1jk$, $j,
k=0, 1$.

(C) Particles 1, 3 and 5 play symmetric and equal role. That is, Alice can apply one unitary operator in
Eq.(\ref{operation1}) on each of one particles $m$ of her particles 1, 3 and 5, and one in Eq.(\ref{operation2})
on each of another particles $n$ of her particles 1, 3 and 5, and Alice and Bob agree on local operations
$\sigma_{ij}$ and $\sigma_k$ encoded as $ij$ and $k$, respectively; or Alice performs on each of particles $m$
of particles 1, 3 and 5 one unitary operation in Eq.(\ref{operation1}), and on each of particles $n$ ($n\neq m$)
of particles 1, 3 and 5 one unitary operation in Eq.(\ref{m21}), and Alice and Bob encode $\sigma_{ij}$ as $ij$,
and encode $I$ and ${\rm i}\sigma_y$  as 0 and 1, respectively. Here $i, j, k=0, 1$.

(D) Alice can also apply one operator in Eq.(\ref{m21}) on each of her particles 1, 3 and 5, and she and Bob
agree beforehand as the  encoding of Eq.(\ref{m22}).

3. Security

Our QSDC protocol is based on EPR pairs, so the proof of security is similar to those in Refs. [3,  25, 34, 35].
   Once the security of the quantum channel is assured, which means that an eavesdropper Eve has not acquire
particles 2, 4 and 6, i.e. Alice and Bob shares pure EPR pairs (perfect quantum channel), then no information is
leaked to Eve. Hence
 our proposed protocol is secure, even if the shared quantum channels are
public.

In summary, we present a new deterministic secure method for direct communication, where the two spatially
separated parties  faithfully transmit secret messages using entanglement swapping  and detect eavesdroppers by
the correlations of entanglement swapping results. A set of ordered Einstein-Podolsky-Rosen (EPR) pairs is used
as  a quantum information channel for sending secret messages directly. After insuring the safety of the quantum
channel, the sender Alice encodes the secret messages directly by applying a series local operations on her
particles sequence according to their stipulation and  send them to  distant receiver Bob.
 Using three EPR pairs, three bits of secret classical information can
be faithfully transmitted from Alice to remote Bob without revealing any information to a potential
eavesdropper.
  By both Alice and Bob's
GHZ state measurement results, Bob is able to read out the encoded secret messages directly.  The protocol is
completely secure if perfect quantum channel is used, since there is not a transmission of the qubits
 carrying the secret message between Alice and Bob in the public channel. This scheme is also a
  quantum key distribution (QKD) scheme, in which via three EPR pairs, six bits of secret key (three certain bits
  and three random bits) can be shared between two spatially separated parties that play symmetric and equal
  roles.\\

{\noindent\bf Acknowledgments}\\

{\noindent This work was supported by National Natural Science Foundation of China under Grant No.
10271081 and Hebei Natural Science Foundation under Grant No. A2004000141.}\\

{\parindent=0cm \bf References } \footnotesize
\begin{tabbing}
  aaaa \= a  \kill
  $[1]$ \> C. H. Bennett and G. Brassard, in {\it Proceedings of the IEEE International Conference on Computers,
   Systems and Signal
       Processing},\\~~~~~~ Bangalore, India (IEEE, New York, 1984), pp.175-179.\\
  $[2]$ \> A. K. Ekert, Phys. Rev. Lett. {\bf  67}, 661 (1991). \\
  $[3]$ \> C. H. Bennett, G. Brassard, and N. D. Mermin,  {\it Phys. Rev. Lett.} {\bf  68},  557 (1992) \\
  $[4]$ \> C. H. Bennett,  {\it Phys. Rev. Lett.} {\bf  68},  3121 (1992). \\
  $[5]$ \> C. H. Bennett and S. J. Wiesner,   {\it Phys. Rev. Lett.} {\bf 69},  2881 (1992). \\
  $[6]$ \> L. Goldenberg and L. Vaidman,   {\it Phys. Rev. Lett.} {\bf 75},  1239 (1995). \\
  $[7]$ \> B. Huttner, N. Imoto, N. Gisin, and T. Mor,  {\it  Phys. Rev.  A} {\bf 51},  1863 (1995). \\
  $[8]$ \> M. Koashi and N. Imoto,  {\it Phys. Rev. Lett.} {\bf 79},  2383 (1997). \\
  $[9]$ \> D. Bru$\ss$,  {\it Phys. Rev. Lett.} {\bf 81},  3018 (1998). \\
  $[10]$ \>  W. Y. Hwang, I. G. Koh,  and Y. D. Han,  {\it Phys.  Lett. A} {\bf 244},  489 (1998). \\
  $[11]$ \> A. Cabello, {\it  Phys. Rev. Lett.} {\bf 85},  5635  ( 2000).\\
  $[12]$ \> A.  Cabello,  {\it Phys. Rev. A}  {\bf 61},  052312 (2000).\\
  $[13]$ \> G. L. Long and X. S. Liu,  {\it Phys. Rev. A} {\bf 65},  032302 (2002).\\
$[14]$ \>  P. Xue, C. F. Li, and G. C. Guo,  {\it  Phys. Rev. A} {\bf 65},  022317 (2002).\\
$[15]$ \> S. J. D. Phoenix, S. M. Barnett, P. D. Townsend, and  K. J. Blow, {\it J. Modern Optics} {\bf 42}, 1155 (1995).\\
$[16]$ \> H. -K. Lo, H. F. Chan, and M.  Ardehali, arXiv: quant-ph/0011056.\\
$[17]$ \> D. Song, {\it Phys. Rev. A} {\bf 69},  034301 (2004).\\
$[18]$ \> X. B. Wang, {\it Phys. Rev. Lett.} {\bf 92},  077902 (2004).\\
$[19]$ \> C. Li, H. S. Song, L. Zhou and C. F. Wu, J. of Opt. B: Quantum and semi-classical optics {\bf 5}, 115 (2003).\\
$[20]$ \> C. Li, H. S. Song, L. Zhou and C. F. Wu, arXiv: Quant-ph/0310077.\\
$[19]$ \> M. Hillery, V. Bu\v{z}ek, and A. Berthiaume, {\it Phys. Rev. A} {\bf 59}, 1829 (1999).\\
$[20]$ \> K. Shimizu and N. Imoto, {\it Phys. Rev. A} {\bf 60}, 157 (1999).\\
$[21]$ \> K. Shimizu and N. Imoto, {\it  Phys. Rev. A} {\bf 62}, 054303 (2000).\\
$[22]$ \> A. Beige {\it et al},  {\it Acta Phys. Pol. A} {\bf 101},  357 (2002).\\
$[23]$ \> K. Bostr\"{o}m and T. Felbinger, {\it Phys. Rev. Lett.} {\bf 89},  187902  (2002).\\
$[24]$ \> A. W$\acute{\rm o}$jcik, {\it Phys. Rev. Lett.} {\bf 90},  157901  (2003).\\
$[25]$ \> F. G. Deng, G. L. Long, and X. S. Liu, {\it Phys. Rev. A}  {\bf 68}, 042317 ( 2003).\\
$[26]$ \> C. H. Bennett {\it et al}, {\it Phys. Rev. Lett.} {\bf 70},  1895  (1993).\\
$[27]$ \> F. L. Yan and X. Q. Zhang,  arXiv: quant-ph/0311132.\\
$[28]$ \> A. Karlsson and M. Bourennane, {\it Phys. Rev. A} {\bf 58},  4394 (1998). \\
$[29]$ \> T. Gao, arXiv: quant-ph/0312004.\\
 $[30]$ \> T. Gao, F. L. Yan, and Z. X. Wang, arXiv: quant-ph/0403155.\\
$[31]$ \> M. Zukowski, A. Zeilinger, M. A. Horne, and A. K. Ekert, {\it Phys. Rev. Lett.} {\bf 71}, 4287 (1993).\\
$[34]$ \> H. Inamori, L. Rallan, and V. Verdral, J. Phys. A {\bf 34}, 6913 (2001)\\
$[35]$ \> E. Waks, A. Zeevi, and Y. Yamamoto, Phys. Rev. A {\bf 65}, 052310 (2002).
\end{tabbing}

\end{document}